# Simulation of the propagation of a cylindrical shear wave :

# non-linear and dissipative modelling

D. Jeambrun (*), Y. Moreau (*), J.L.Costaz(**), J.P. Tourret(**),
P. Jouanna(***), G. Lecoy(*).

(*) Centre d'Électronique de Montpellier, Montpellier, France.
(**) Électricité de France Service Études Production Thermique Et Nucléaire, Lyon, France.
(***) Dynamique et Thermodynamique des Milieux Complexes, Université Montpellier II, Montpellier, France.

*Abstract*

The simulation of a wave propagation caused by seismic stimulation allows to study the behaviour of the environment and to evaluate the consequences. The model involves the wave equation with a hysteresis loop in the stress-strain relationship. This induces non-linearities and, at the vertices of the loop, non-differentiable mathematical operators. This paper offers a numerical process which works out this simulation.

**Introduction**

Protection against earthquakes is an increasingly important objective for civil engineering, especially for those buildings in areas of high seismic risk [1][2]. Of course, the problem is to know how a structure is going to respond under earthquake loading. But, before seismic motion reaches the structure, it must pass through the various soil layers underlying the site. Depending on the behaviour of the soils, the structure may be quite safe, or doomed to collapse. For this reason, the engineers look with interest upon the problem of soil response under dynamic loading.

The soil characterisation needs an in situ experimental device. The elaboration of the device requires a theoretical study before it is manufactured, and some parameters seem hard to evaluate directly from the measures. This is especially the case when considering the thermal dissipation. A link between experimental results and simulation results should allow us to identify the values of soil parameters. Furthermore, an experimentation with a single well for both stimulation and response may be designed [3], much cheaper than the usual cross holes system.

The simulation is based on the propagation of cylindrical shear waves. This configuration corresponds to the experimental system project. Several simulations have been performed through linear models which consider the soil as elastic. A linear model requires the determination of the shear modulus $G_0$ as well as the soil density $\rho$ to compute the propagation of the shear wave. In the real case of important seismic stress, elastic behaviours can no longer be assumed.

The first improvement is to consider a non-linear phenomenon, introducing an internal damping parameter α (saturation effect of the stress for the high level of strain) added to the geometrical damping (dispersal of the energy density when the distance from the source is increased; obviously due to the volume augmentation in correspondence with the distance from the source). This is described by a pure non-linear model.

Moreover, when the strain is important, a part of mechanical energy is converted into thermal energy. This induces a delay on the propagation signals. In one space dimension, this can be modelled by a hysteresis cycle between the stress and the strain (Figure 1).

Simple non-linear modelling enables to adjust parameters, used with dissipative models. The difficulty is to obtain the computation stability and convergence in spite of acceleration discontinuities. These discontinuities follow the non-differentiable mathematical operators due to the elasto-plastic nature of the non-linear dissipative model [4].

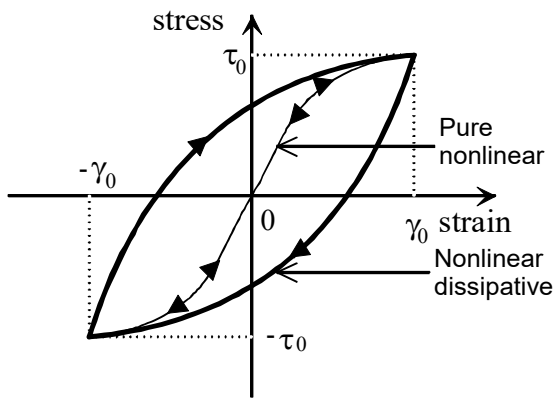
Figure 1 : stress-strain diagrams

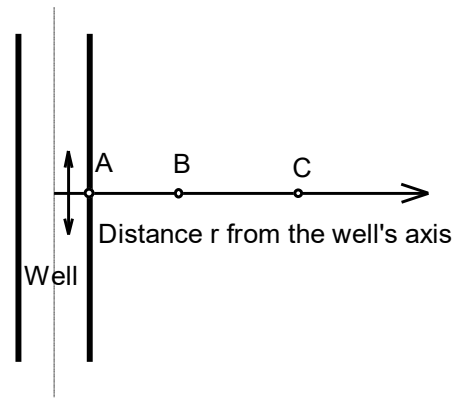
Figure 2 : Test configuration

**Configuration, Assumptions and Models**

Figure 2 shows, schematically, the experimental device (designed for Electricité de France [3]). One expects this device will allow the measures of the displacement and force at the well edge (A). The experimental system might also give the soil accelerations in the well vicinity (B and C).

The soil is assumed to be homogenous and isotropic. The probe is assumed to have an infinite length. Therefore, the wave propagation can be considered as radial and the displacement purely vertical, enabling the use of one-dimensional model.

The macroscopic soil models are based on the momentum balance relationship. In one radial dimension, it is described by the equation (1) where $\tau(r,t)$ is the vertical shear stress in the soil, $w(r,t)$ is the vertical material displacement of the soil, r is the horizontal distance from the axis, t is the time and $\rho$ is the specific mass of the soil. To complete the models, behaviour laws are added to this equation. The simplest is the elastic behaviour (2) where $\gamma(r,t)$ is the vertical shear strain of the soil and $G_0$ is the shear wave modulus at low strain. This behaviour is, in general, valid for a strain less than $10^{-5}$ and allows the computation with the well-know analytic solutions [5].

$$\frac{\partial \tau(r,t)}{\partial r} + \frac{\tau(r,t)}{r} - \rho \frac{\partial^2 w(r,t)}{\partial t^2} = 0 \qquad (1)$$

$$\tau(r,t) = G_0 \gamma(r,t) \tag{2}$$

$$\tau(r,t) = \frac{G_0 \gamma(r,t)}{1 + \delta\alpha\gamma(r,t)} \tag{3}$$

$$\tau(r,t) = \frac{(2G_0 - \alpha\tau_0(r))(\gamma(r,t) + \beta\gamma_0(r)) - 2\beta\tau_0(r)}{2 + \alpha\beta(\gamma(r,t) + \beta\gamma_0(r))} \tag{4}$$

For a higher strain, and before irreversible effects (which occur when the strain is approximately greater than $10^{-4}$), the damping phenomenon must be taken into account. The non-linear behaviour law of Hardin and Drewich [6, 7] (3) (drawn with thin lines in Figure 1), where $\alpha$ is the damping parameter, has this particularity. Moreover, to introduce the dissipative effect, a non-linear and hysteresis cycle behaviour law (4) [2] (drawn with thick lines in Figure 1) can be used. The loop surface is proportional to the thermal energy dispersed in soil. The law (4) depends on two soil parameters: the linear parameter $G_0$ and the dissipative parameter $\alpha$. A $\beta$ coefficient makes possible the loading when $\beta = 1$ and unloading when $\beta = -1$. $\tau_0$ and $\gamma_0$ are the values of the vertex of the hysteresis loop (seen in Figure 1). For the determination of the last three parameters $\beta$, $\tau_0$ and $\gamma_0$ we use a pre-computation with the law (3). The law (4) induces discontinuities at the times when the stress reaches the vertices of the hysteresis loop [4].

**Method of Integration, Stability and Convergence**

To achieve the computation, conditions must be found on the space domain boundary. The displacement w at the well edge is the system input. Far away from the well, an absorbing boundary condition (5) [8] is used, where $r_b$ is the distance from the well axis and $c_s$ is the shear wave celerity. The boundary must be in the elastic behaviour area, in this study the boundary was taken at 10 m.

$$\frac{\partial w(r_b,t)}{\partial r} + \frac{1}{c_s}\frac{\partial w(r_b,t)}{\partial t} + \frac{w(r_b,t)}{2r_b} = 0 \tag{5}$$

The non-linearity and the dissipative effect are major near the well. The space steps must be small in this zone, but the computation must be made far from the well to be allowed to use the absorbing boundary equation (5). Therefore, the finite difference method with variable steps is used for space. With this method, space steps can be thin in the well vicinity, whereas, far from the well, the space steps can be larger. Here the space steps are equal to 0.1 mm at the well edge with a 5% progression at every step.

The time evolution is insured by a Newmark-Wilson method [9] [10]. This method allows the stabilisation of the computation when the Wilson parameter is greater than one. The simplest methods tested are unstable with this non-linear dissipative model. The instabilities are caused by the discontinuities of the acceleration.

At each time step, the convergence of the non-linear equations is achieved with the help of the classic Newton-Raphson process. The Newton-Raphson iteration itself requires successive solutions of linear tri-diagonal systems.

The Newmark parameters a and b are taken such as $a = b = 1$. Although this condition is more restrictive than the stability criterion in the linear case, we didn't note the overall stability in the cases we tried. The Wilson parameter θ behaves as a low pass frequency filter. Therefore, when the space steps are small enough, this parameter ensures the stability in the well vicinity in spite of the acceleration discontinuities. The display of the acceleration close to the well allows the checking of the stability.

When the non-linear parameter α is equal to zero, the comparison with the linear analytic computation makes errors less than 1%. The pure non-linear model is computed in the first half of the periods, the non-linear dissipative model is used in the last half of the periods.

**Results and Discussion**

We chose the soil values $\rho = 2000$ kg/m3, $G_o = 7.2\ 10^7$ Pa and $\alpha = 2000$ - unless otherwise marked -, the experimental values $r_{well} = 0.1$ m, $T = 0.1$ s, $|w_{well}| = 0.1$ mm and the computation parameters $\Delta t = 1.6$ ms, $\Delta r = 0.1$ m at the well edge with a 5% progression, $r_b = 10.1$ m. The Newmark-Wilson parameters are $a = b = 1$ and $\theta = 1.1$.

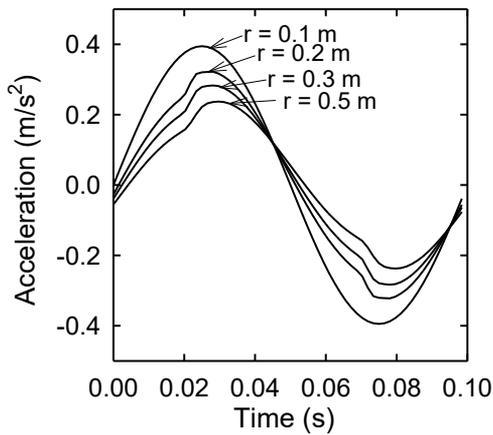

Figure 3 : Acceleration vs. distance

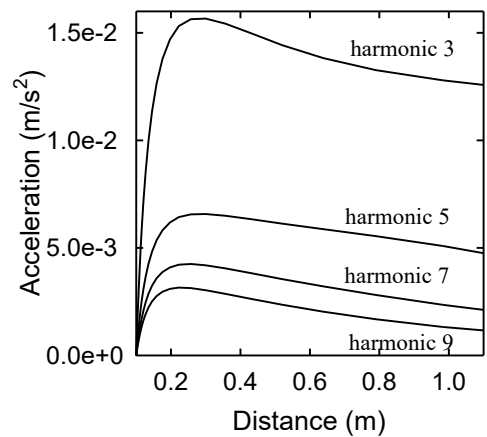

Figure 4 : Harmonic modules vs. distance

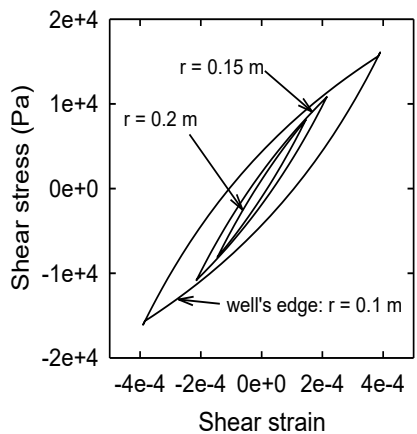

Figure 5 : Hysteretic loop vs. various distances

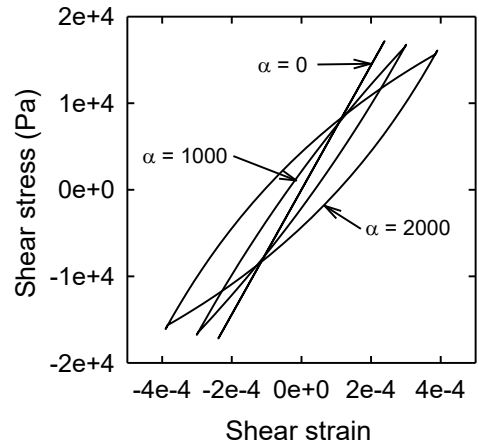

Figure 6 : Hysteretic loop for various α

Figure 3 shows the acceleration at the well edge (r = 0.1 m) and in the well vicinity at r = 0.2 m, r = 0.3 m and r = 0.5 m from the axis. The discontinuities can be seen in the last three distances. The harmonic modules versus the distance from the well edge are shown in figure 4. The even harmonics are negligible. One sees that the harmonics are generated in the first 20 cm. The figures 5 and 6 show hysteretic cycles respectively at different distances from the well axis when $\alpha$ = 2000 and for some other values of $\alpha$. This shows the decrease of the loop's surface with respect to the distance from the source and with respect to the $\alpha$ decrease.

To have a fine knowledge of the mechanic properties of soils, civil engineers need to identify the parameters $G_0$ and $\alpha$. The elasticity parameter $G_0$ has a well-known influence on the measurable variables. As soon as the stress is strong enough and before the soil ruptures, previous works [1] and this paper show that the plasticity parameter $\alpha$ also influences these variables. It can be checked by the generation of the odd harmonics on accelerations within the non-linear area close to the well versus $\alpha$. Thus, two kinds of methods could be applied for this parameter identification with in situ measures. On the one hand, both parameters can be found by a disconsolation process such as the non-linear chi-squared minimisation. On the other hand, the $G_0$ parameter can be worked out with low stress measures where the influence of $\alpha$ is negligible. When $G_0$ is known, $\alpha$ can be identified with the help of the computed odd harmonic accelerations.

**Conclusions**

The computer code based on pure non-linear elastic and non-linear dissipative models using adapted algorithms, time integration of Newmark-Wilson, spatial mesh with variable steps and a Newton-Raphson process for the non-linear computation, overcomes the stability difficulties related to acceleration discontinuities present in the case of high seismic stress. The influence of the dissipative model has been shown on the acceleration spectrum and on the hysteresis loop. These simulations would allow the identification of soil parameters by comparison with in situ measures.